This item is the peer-reviewed, accepted manuscript of:

with the same content as the published version but without the editorial typesetting and formatting. To refer to or to cite this paper, please use the citation to the published version.

# Chaotic dynamics of granules-beam coupled vibration: Route and threshold

Hang Li [1], Jian Li [1,*], Hongzhu Fei [1], Guangyang Hong [1,*], Jinlu Dong [1], Aibing Yu [2,3]

[1] Key Laboratory of Structural Dynamics of Liaoning Province, College of Sciences, Northeastern University, Shenyang, 110819, China.

[2] ARC Research Hub for Computational Particle Technology, Department of Chemical Engineering, Monash University, Clayton, VIC 3800, Australia.

[3] Centre for Simulation and Modelling of Particulate Systems, Monash-Southeast University Joint Research Institute, Suzhou, China.

[*] Corresponding author. Email: jianli@mail.neu.edu.cn (J. Li), 1113125092@qq.com (G. Hong).

## Abstract

Although granular materials are the second most processed in industry after water, the theoretical study of granules-structure interactions is not as advanced as that of fluid-structure interactions due to the lack of a unified view of the constitutive relation of granular materials. In the previous work [1], the theoretical model of granules-beam coupled vibration was developed and verified by experiments. However, it was also found that the system exhibits significant stiffness-softening Duffing characteristics even under micro-vibration, which implies that chaotic responses may be reached under certain conditions. To reveal the route and critical conditions for the system to enter chaos, in the present work, the chaotic dynamics of the system are studied. In qualitative analysis, Melnikov method is applied to analyze the instability behavior of the perturbed heteroclinic orbit of the system, thus the critical condition for the system to enter Smale horseshoe chaos, i.e., the Melnikov criterion, is obtained. The validity of the criterion is verified numerically. In experimental studies, the existence of chaotic responses and the route to chaos for granules-beam coupled vibrations is revealed. The experimental results suggest that the system response first experiences symmetry-breaking then undergoes a complete period-doubling cascade, and finally enters single-scroll chaos. In addition, although the Additional Dynamic Load (ADL) generated by granular media is highly complicated, a general and simple evolution pattern of the chaos threshold is found by parametric experiments, which is also supported by the Melnikov criterion. In short, the chaotic dynamics of granules-beam coupled vibration is revealed, which is a contribution to the engineering vibration aspect. On the theoretical side, an impressive result lies in that for the first time, the Melnikov criterion of such fractional-order systems is obtained in a global and closed form, which provides, respectively, improvement suggestions and reference results for the existing and future research on chaotic dynamics of fractional-order systems, especially of the Duffing-type systems.

vibration; Applied fractional dynamics; Chaos threshold; Melnikov method for fractional-order systems.

# 1 Introduction

The dynamic characteristics of engineering structures in media are significantly different from those in vacuum. The present work is dedicated to the study of granules-structure coupled vibrations. To facilitate the introduction, it seems appropriate to briefly review the research on fluid-structure ones.

Research on the fluid-structure coupled vibrations can be traced back to Rayleigh's work [2] on waves and acoustics in 1877, when he calculated the increment of inertia of a rigid circular plate vibrating in an inviscid and incompressible liquid. As for the elastic plate, this was the problem studied by Lamb [3], who found that the natural frequencies of the plate immersed in water are generally lower than those in vacuum. A vivid explanation for the changes in these dynamic characteristics of structures can be found in the correspondence [4] between Kelvin and Stokes — "There will be an exceedingly thin stratum of fluid round the solid through which the velocity of the water varies continuously from the velocity of the solid to the velocity in the solution for inviscid fluid...; The less the viscosity, the thinner is this layer…". Nowadays, benefiting from the development in fluid mechanics, theoretical approaches to study fluid-structure interactions are well developed, and various problems of coupled vibration have been theoretically studied on this basis, such as the beams immersed in liquid [5-7], the plates immersed in liquid [8-10], and the pipes conveying fluids [11-13]. In general, the main purpose of fluid-structure coupled vibration research is to evaluate the structural response and the fluid flow as well as their mutual influence. Almost for the same purpose, granular materials (the second most processed industrial material after water [14]) and structure coupled vibrations have started to be of interest in recent years [15-24].

However, in granular mechanics, there is currently no theoretical framework that reaches the level of the Navier-Stokes equations in fluid mechanics, which is a major obstacle for theoretically studying the granules-structure coupled vibrations. Specifically, the macroscopic mechanical behavior of granular materials may exhibit as solid-like materials that can withstand certain loads [15, 25, 26]; or, due to external driving, they may exhibit as liquid-like or gas-like materials that can flow over a large range [27-31]. For a long time, engineers have tended to model granular matter based on continuum mechanics [15, 32], while physicists are primarily interested in using the methods of statistical mechanics and fluid dynamics, which usually lead to phenomenological models, such as the $\mu(I)$ rheology of granular flow [30, 33]. To date, there is still a lack of a unified view on the constitutive relation of granular matter. Reviewing the existing research on the granules-structure coupled vibrations [17-24], one can learned that benefiting from the enhancement of computational capability, numerical methods, such as discrete element method, are usually combined with experimental study, and some meaningful conclusions and insights are obtained, mainly in the sense of statistical mechanics.

For the above reasons, no constitutive model of granular matter was considered in the previous work [1]. Instead, inspired by the research on fluid-structure interactions [5-13] and inerter-type vibration control techniques [34-42], granules-beam coupled vibration was theoretically modeled from the perspective of mechanical system dynamics. However, it was also found experimentally that the system exhibits significant stiffness-softening Duffing characteristics even under micro-vibration, which implies that the system response may reach chaos under certain conditions. From the perspective of dynamic design, the intrinsic pseudo-randomness of chaos will compromise the optimality of designs based on periodic solutions, such as designs based on the $H_\infty$ optimization criterion [37, 38] and the $H_2$ optimization criterion [43]. In addition, chaotic signals are promising for applications in weak signal detection [44] and machinery fault diagnosis [45, 46]. For the above purposes, i.e., to suppress chaos and to utilize chaos, chaotic dynamics has been an intriguing topic in the study of mechanical system dynamics [47, 48]. The questions that naturally arise are how the system evolves from periodic to chaotic vibration and what the critical parameters of the system are. Motivated by these two questions, the present work aims to reveal the route to chaos of the granules-beam coupled vibrations and to determine the corresponding chaos threshold analytically.

First, Melnikov method [49] is applied to analyze the instability behavior of the perturbed heteroclinic saddle-saddle connection in the phase space. Subsequently, the Melnikov criterion for the system is obtained, which serves as an analytical tool for determining the critical parameters that cause the system response to enter chaos in the sense of the Smale horseshoes. Then, to verify the Melnikov criterion and to reveal the route to chaos, simulation studies are carried out based on the theoretical model. Then, the validity of the Melnikov criterion is confirmed by the agreement between the reference values and the theoretical values. In the experimental study, the results show the single-scroll chaotic attractor of the system and the period-doubling route to such chaos. Prior to this bifurcation, the system response first undergoes symmetry-breaking of the periodic orbit. It is also impressive that these experimental results are in good agreement with those based on the theoretical model. In addition, the Melnikov criterion reveals a general and simple evolution pattern of the chaos threshold, which is further well verified by the experimental study. In short, the mechanism of chaotic vibrations of granules-beam coupled systems is revealed theoretically and verified experimentally, which is one of the main contributions in this work.

On the other hand, since the introduction of the Melnikov method into fractional dynamics [50], the Melnikov criteria of fractional-order systems have not yet been obtained in explicit form but rather depend on local analytical [51-57] or numerical methods [58-63]. Here, the Melnikov criterion of such fractional-order system is derived for the first time in a global and closed form. From the theoretical perspective, this result is not only applicable to the system studied in the current work, but also can be considered as an improvement suggestion and reference result for the existing and future analytical research on the chaotic dynamics of fractional-order systems, especially the Duffing-type ones. This is also one of the main contributions in this work.

The rest of this paper is organized as follows. In Section 2, the governing equation of the granules-beam coupled vibration developed in the previous work is briefly introduced. In Section 3, the global dynamics analysis on the system is performed based on Melnikov method, and the exact Melnikov criterion under first-order perturbation is derived through rigorous analysis. In Section 4, the validity of the Melnikov criterion is verified numerically. In Section 5, the existence of chaotic responses in granules-beam coupled vibration and the route to chaos are revealed experimentally, and the evolution pattern of the chaos threshold is investigated by combining experiment and theory. In Section 6, the main conclusions and findings of this work are summarized, and finally recommendations for future research on this topic are presented.

## 2 Mathematical model

Consider a slender beam buried at depth $H$ in a dense granular media, as illustrated schematically in Fig. 1. The granular media is composed of monodisperse granules with diameter $d_0$. The modulus of elasticity, second moment of area, volume density, and cross-sectional area of the beam are denoted by $E$, $I$, $\rho$, and $S$, respectively. The geometric sizes of the beam are $L\,(\text{length})\times T_h\,(\text{thickness})\times b\,(\text{width})$, and a local coordinate system $(x, y, z)$, which is along $L$, $T_h$, and $b$, is established on the beam. The beam is nominally horizontal, and the effect of gravity on the beam is neglected here. To study the characteristics of granules-beam coupled vibration, an external force $F_d(x,t)$ is used to is excite the beam. At the axial position $x$ and time $t$, the lateral vibration of the beam is specified by the deflection $w(x,t)$ of the neutral plane.

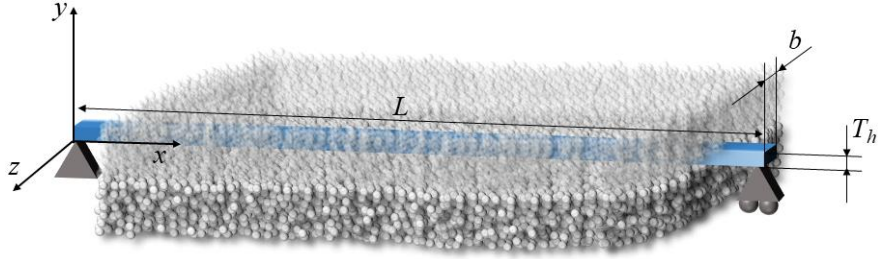

**Fig. 1.** Schematic representation of the granules-beam coupled vibration (reprinted from the previous work [1] under license from Elsevier).

It can be expected that the dynamic characteristics of the beam will be significantly changed due to the ADL provided by the granular media. Also, the physical properties of the granular media will be affected due to the beam vibration, which in turn affects the ADL. As presented in the Introduction, such coupled vibrations were theoretically modeled in the previous work [1], and the developed model was validated by numerous experiments. In the current work, the focus is on the chaotic dynamics of granules-beam coupled systems.

The developed theoretical model of the granules-beam coupled vibration is described as follows.

$$\rho S \frac{\partial^2 w}{\partial t^2} + EI \frac{\partial^4 w}{\partial x^4} + \eta I \frac{\partial^5 w}{\partial x^4 \partial t} + k_L w + k_N w^3 + \beta_0 \frac{\partial^q w}{\partial t^q} = F_d(x,t) \qquad (1)$$

where $\eta$ is the viscosity coefficient originated from the Kelvin-Voigt viscoelastic model for modeling the structural damping of the beam. $k_L$ ( $k_L > 0$ ) and $k_N$ ( $k_N < 0$ ) quantify the elastic restoring force provided by the solid-like phase granular media on the beam. $\beta_0$ ( $\beta_0 > 0$ ) quantifies the inertia-viscous effect provided by the fluidized granular media, which is modeled by the following $q$ -order ($1 < q < 2$) left Caputo derivative [64].

$$\frac{\partial^q w(x,t)}{\partial t^q} = \frac{1}{\Gamma(2-q)} \int_\alpha^t (t-s)^{1-q} \frac{\partial^2 w(x,s)}{\partial s^2} ds \qquad (2)$$

where $\Gamma(\bullet)$ stands for Euler Gamma function.

The following simply-simply supported boundary condition and centralized harmonic excitation are considered here.

$$w(0,t) = 0, \quad w(L,t) = 0, \quad \frac{\partial^2 w(x,t)}{\partial x^2}\Big|_{x=0} = 0, \quad \frac{\partial^2 w(x,t)}{\partial x^2}\Big|_{x=L} = 0 \tag{3a}$$

$$F_d(x,t) = \delta(x - \xi) F_0 \sin\left[(2\pi f)t\right] \tag{3b}$$

where $\delta(\bullet)$ stands for the Dirac function, $\xi$ is the $x$-axis coordinate of the excitation position. $F_0$ and $f$ denote the excitation amplitude and frequency, respectively.

Next, chaotic dynamics of granules-beam coupled vibration is revealed based on the global dynamics analysis for the above theoretical model and the corresponding experimental study. It is worth highlighting that no modifications are further introduced to the theoretical model here. In other words, it is valid not only for the periodic vibration but also for the chaotic one.

## 3 Global dynamics

### 3.1 Perturbation analysis

Herein, Melnikov method is employed to analyze the chaotic dynamics of granules-beam coupled vibration. According to Nayfeh et al. [65], in damped systems, the modes that are not directly excited by external source or indirectly excited by internal resonance will eventually decay over time. Therefore, for the common case where $n$-th primary resonance exists without active internal resonance, Eq. (1) can be discretized by the following single-mode Galerkin discretization.

$$w(x,t) = \Phi_n(x) U_n(t) \tag{4}$$

where the $n$-th modal function $\Phi_n(x) = \sin\dfrac{n\pi x}{L}$, $n \in \mathbb{Z}^+$.

Then, substituting Eq. (4) into Eq. (1), multiplying the obtained equation by $\Phi_n(x)$, and then integrating the product from $0$ to $L$ yields the governing equation of the $n$-th primary resonance as follows.

$$\frac{\rho SL}{2}\ddot{U}_n + \frac{\eta In^4\pi^4}{2L^3}\dot{U}_n + \left(\frac{EIn^4\pi^4}{2L^3} + \frac{Lk_L}{2} + \frac{3Lk_N}{8}U_n^2\right)U_n + \frac{\beta_0 L}{2}\mathrm{D}_t^q(U_n) = F_0\sin\frac{n\pi\xi}{L}\sin\left[(2\pi f)t\right] \quad (5)$$

where $\mathrm{D}_t^q(\bullet)$ stands for the $q$-order left Caputo derivative operator $\dfrac{\mathrm{d}^q}{\mathrm{d}t^q}(\bullet)$.

Note that the Galerkin discretization above is based on the synchronous vibration assumption that all generalized coordinates simultaneously pass through their amplitude and zero points of potential energy, even in the chaotic response. In other words, the chaotic response of that coupled vibration system is in the time domain, while independent of spatial position. In view of the micro-vibration behavior, the material homogeneity of the beam, and the symmetry of the macroscopic behavior in the granular media [66], this assumption is allowed to be employed. In fact, the assumption is supported, and the corresponding experimental evidence will be given in Section 5.

Further, introducing the state variable $\boldsymbol{u} = \left[u_1, u_2\right]^{\mathrm{T}} = \left[U_n, \dot{U}_n\right]^{\mathrm{T}}$ and the following parametric transformations to simplify the symbolic expression for Eq. (5).

$$m = \frac{\rho SL}{2}, \quad c = \frac{\eta In^4\pi^4}{2L^3}, \quad k = \frac{EIn^4\pi^4 + k_L L^4}{2L^3},$$

$$\beta_1 = \frac{\beta_0 L}{2}, \quad F_1 = F_0\sin\frac{n\pi\xi}{L}, \quad \omega = 2\pi f, \quad \gamma_1 = \frac{3Lk_N}{8}, \quad (6)$$

$$\omega_0^2 = \frac{k}{m}, \quad \gamma = -\frac{\gamma_1}{m}, \quad \varepsilon\beta = \frac{\beta_1}{m}, \quad \varepsilon F = \frac{F_1}{m}, \quad 2\varepsilon\mu = \frac{c}{m},$$

where the dimensionless bookkeeping parameter $\varepsilon$ is introduced just for Melnikov analysis. No further modifications or restrictions are introduced to the model here. In other words, the model is valid not only for the periodic vibration, but also for the chaotic vibration.

Thus, Eq. (5) is rewritten as the following form.

$$\dot{\boldsymbol{u}} = g(\boldsymbol{u}) + \varepsilon P(\boldsymbol{u},t) = \begin{bmatrix} u_2 \\ -\omega_0^2 u_1 + \gamma u_1^3 \end{bmatrix} + \varepsilon\begin{bmatrix} 0 \\ F\sin\omega t - 2\mu u_2 - \beta\mathrm{D}_t^q(u_1) \end{bmatrix} \quad (7)$$

The corresponding unperturbed system of Eq. (7) is given by

$$\dot{\boldsymbol{u}} = g(\boldsymbol{u}) \tag{8}$$

with three fixed points at $C:(0,0)$, $S_1:\left(-\dfrac{\omega_0}{\sqrt{\gamma}},0\right)$ and $S_2:\left(\dfrac{\omega_0}{\sqrt{\gamma}},0\right)$, where $C$ is the center equilibrium, while $S_1$ and $S_2$ are saddle points.

As illustrated in Fig. 2, the phase portrait of the unperturbed system Eq. (8), which is simulated by the fourth-order Runge-Kutta method, has a heteroclinic orbit $\boldsymbol{u}^0$ that connects $S_1$ and $S_2$. By integrating Eq. (8) while substituting the coordinates of $S_1$ and $S_2$, one can obtain the solution of the heteroclinic orbit $\boldsymbol{u}^0$ as follows.

$$\boldsymbol{u}_\pm^0(t) = \begin{bmatrix} \pm\dfrac{\omega_0}{\sqrt{\gamma}}\tanh\left(\dfrac{\omega_0 t}{\sqrt{2}}\right) \\ \pm\dfrac{\omega_0^2}{\sqrt{2}\gamma}\operatorname{sech}^2\left(\dfrac{\omega_0 t}{\sqrt{2}}\right) \end{bmatrix} \tag{9}$$

For the perturbed system Eq. (7), the stable manifold $W^s$ and the unstable manifold $W^u$ near the saddle points split due to the perturbation $P(\boldsymbol{u},t)$. Consequently, they may intersect transversally at a certain critical perturbation. According to the Smale-Birkhoff theorem [67], such a transversal intersection point serves as an indicator that the perturbed system enters Smale horseshoe chaos. Furthermore, by measuring the distance between $W^s$ and $W^u$, the existence of transverse homoclinic points can be detected.

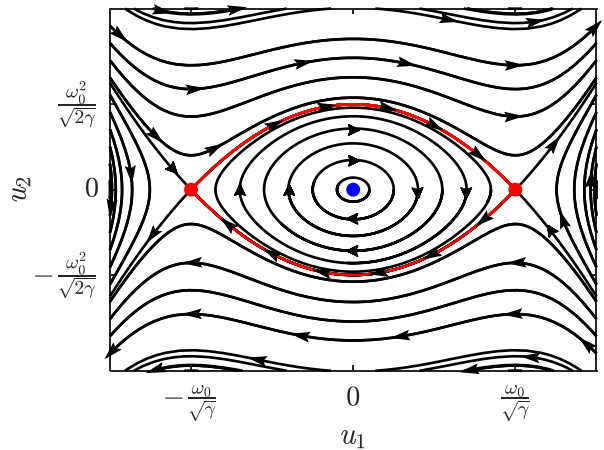

**Fig. 2.** Phase portrait of the unperturbed system Eq. (8), where the heteroclinic orbit $\boldsymbol{u}^0$ and its two saddle points $S_1$ and $S_2$ are highlighted in red, and the center equilibrium $C$ are highlighted in blue.

## 3.2 Exact Melnikov criterion under first-order perturbation

According to Melnikov method, the distance between $\mathrm{W}^u$ and $\mathrm{W}^s$ near $\boldsymbol{u}_+^0$ (similar for $\boldsymbol{u}_-^0$) after splitting can be evaluated by the following Melnikov function $M(\tau)$.

$$
\begin{aligned}
M(\tau) &= \int_{-\infty}^{+\infty} g\left[\boldsymbol{u}_+^0(t)\right] \wedge P\left[\boldsymbol{u}_+^0(t), t+\tau\right] \mathrm{d}t \\
&= \int_{-\infty}^{+\infty} \frac{\omega_0^2}{\sqrt{2}\gamma} \operatorname{sech}^2 \frac{\omega_0 t}{\sqrt{2}}\left[F \sin \omega(t+\tau) - \frac{2\mu\omega_0^2}{\sqrt{2}\gamma} \operatorname{sech}^2 \frac{\omega_0 t}{\sqrt{2}} - \frac{\beta\omega_0}{\sqrt{\gamma}} \mathrm{D}_t^q \tanh \frac{\omega_0 t}{\sqrt{2}}\right] \mathrm{d}t \quad (10)\\
&= I_1 + I_2
\end{aligned}
$$

where the operator $\wedge$ denotes the wedge product, and the integrals $I_1$ and $I_2$ are as follows.

$$
\begin{aligned}
I_1 &= \int_{-\infty}^{+\infty} \frac{\omega_0^2}{\sqrt{2}\gamma} \operatorname{sech}^2 \frac{\omega_0 t}{\sqrt{2}}\left[F \sin \omega(t+\tau) - \frac{2\mu\omega_0^2}{\sqrt{2}\gamma} \operatorname{sech}^2 \frac{\omega_0 t}{\sqrt{2}}\right] \mathrm{d}t \\
&= \frac{\sqrt{2}}{3\gamma}\left(3\pi\omega F \sqrt{\gamma} \operatorname{csch} \frac{\pi\omega}{\sqrt{2}\omega_0} \sin \omega\tau - 4\mu\omega_0^3\right)
\end{aligned} \tag{11a}
$$

$$
I_2 = \int_{-\infty}^{+\infty} -\frac{\omega_0^2}{\sqrt{2}\gamma} \operatorname{sech}^2 \frac{\omega_0 t}{\sqrt{2}} \cdot \left(\frac{\beta\omega_0}{\sqrt{\gamma}} \mathrm{D}_t^q \tanh \frac{\omega_0 t}{\sqrt{2}}\right) \mathrm{d}t \tag{11b}
$$

$I_1$ can be easily calculated as given in Eq. (11a). However, for $I_2$, the perturbation of the invariant manifolds by fractional-order elements, it has not yet been obtained globally in explicit form in the existing related works, but rather by local analytical [51-57] or numerical methods [58-63]. In general, by the former, the resulting Melnikov criterion can be obtained in an explicit form although it is constrained by the frequency range of the approximate solutions for nonlinear resonance [1, 68]; while by the latter, the resulting Melnikov criterion is useful for global numerical estimation although it is non-explicit.

In view of the above, the current work is devoted to obtain the Melnikov criterion in a global and closed form based on rigorous analysis. Importantly, to ensure that the perturbations provided by the fractional-order element to orbits lying on unstable manifolds $W^u$ in $(-\infty, \tau)$ are properly calculated, Caputo derivative Eq. (2) should be defined from $\alpha = -\infty$ rather than $\alpha = 0$ as in the existing works. This setup does not conflict with the theoretical analysis on periodic vibrations in the previous work [1], because the governing equation (1) is still allowed to be defined at $t_0 = 0$, which just implies that the initial state of the fractional-order element consists of the entire history in $(-\infty, 0]$.

Therefore, in order to evaluate the Melnikov function of the perturbed system globally and analytically, Eq. (11b) should be integrated as follows.

$$I_2 = \frac{\beta \omega_0^5}{\sqrt{2}\gamma} \frac{1}{\Gamma(2-q)} \int_{-\infty}^{+\infty} \operatorname{sech}^2 \frac{\omega_0 t}{\sqrt{2}} \int_{-\infty}^{t} (t-s)^{1-q} \tanh \frac{\omega_0 s}{\sqrt{2}} \operatorname{sech}^2 \frac{\omega_0 s}{\sqrt{2}} \, \mathrm{d}s \, \mathrm{d}t$$
$$= \frac{\beta \omega_0^5}{\sqrt{2}\gamma} \frac{1}{\Gamma(2-q)} I_3 \tag{12}$$

For convenience, the detailed analysis for $I_3$ is presented in Appendix A. Then, according to Eq. (A8), $I_3$ is finally obtained as follows.

$$I_3 = 2^{\frac{3-q}{2}} \cdot \frac{q(q^2-1)}{\pi^q \omega_0^{3-q}} \cdot \sec \frac{q\pi}{2} \cdot \zeta(q+1) \tag{13}$$

where $\zeta(\bullet)$ stand for Riemann Zeta function.

Finally, Melnikov function $M(\tau)$ is obtained as follows.

$$M(\tau) = \underbrace{\frac{\sqrt{2}\pi \omega F \operatorname{csch} \frac{\pi \omega}{\sqrt{2}\omega_0} \sin \omega \tau}{\sqrt{\gamma}}}_{\text{Part 1}} - \frac{4\sqrt{2}\mu \omega_0^3}{3\gamma} + \underbrace{\frac{2^{1-\frac{q}{2}} \beta \omega_0^{q+2} q(q^2-1)}{\pi^q \gamma} \cdot \frac{\zeta(q+1)}{\Gamma(2-q)} \cdot \sec \frac{q\pi}{2}}_{\text{Part 2}} \tag{14}$$

As discussed previously, $M(\tau)$ measures the distance between the perturbed stable and unstable manifolds, and the intersect transversely between these two, i.e., the existence of a transverse heteroclinic point implies that $M(\tau) = 0$ and $\mathrm{d}M / \mathrm{d}\tau \neq 0$. Impressively, the Part 2 of $M(\tau)$ is always negative within the range $1 < q < 2$. Thus, the existence of a transverse heteroclinic point requires the following condition.

$$\pi\omega F \operatorname{csch} \frac{\pi\omega}{\sqrt{2}\omega_0} > \frac{4\mu\omega_0^3 + 3\beta\pi^{-q}\omega_0^{q+2} \cdot 2^{\frac{1-q}{2}} q\left(1-q^2\right) \cdot \sec\frac{q\pi}{2} \cdot \zeta\left(q+1\right)/\Gamma\left(2-q\right)}{3\sqrt{\gamma}} \quad (15)$$

Unlike the existing results, Eq. (15) is the first closed-form Melnikov criterion of fractional-order Duffing-type systems. The criterion is not only for the system in this paper, but also provides a theoretical reference for the Melnikov analysis on other fractional-order systems. In addition, it is worth noting that there are no unnecessary restrictions on the parametric transformations Eq. (6). Therefore, the criterion can not only serve to estimate the chaos threshold, but also globally characterize the evolution of the chaos threshold, i.e., chaotic vibration is likely to arise under a strong excitation amplitude $F$ and nonlinearity $\gamma$, as well as weak damping $\mu$ and inertia-viscous effects $\beta$.

Overall, the discussion on the existing works and the Melnikov criterion obtained here can be considered as the improvement suggestion and reference result for the existing and future analytical study on the chaotic dynamics of fractional-order systems, especially the Duffing-type ones.

## 4 Numerical validations for Melnikov criterion

Up to this point, Melnikov criterion for granules-beam coupled vibrations is derived. To provide theoretical support for subsequent experimental study, it is necessary to first numerically verify the validity and accuracy of the obtained Melnikov criterion. As discussed in Section 3, the holding of Eq. (15) signifies that the system enters Smale horseshoe chaos. Here, the critical parameter value at which the system first transitions from periodic (or quasi-periodic, if it exists) to chaotic response is defined as the chaos threshold. Next, the theoretical values of the chaos threshold calculated by Melnikov criterion are compared with the reference values obtained by simulation.

Lyapunov exponent, defined by the average divergence or convergence rate of the phase trajectory [69], is a reliable quantitative indicator for determining the type of system response. For the Lyapunov exponent of Eq. (7), it can be computed by formally rewriting Eq. (7) as an autonomous system in the 4-D expanded phase space. The correspondences between the type of response of a 4-D system and the signs of LEs are listed in Table 1. Note that the system is forced to vibrate, so the response cannot be an asymptotically stable equilibrium point, and thus according to Table 1 there is always at least one LE that is 0, called the trivial LE. Therefore, the largest nontrivial LE is the only one required to determine the chaotic response. Here, it is used in combination with the power spectrum to determine the simulation result of the chaos threshold, which is then used as a reference value to compare with the theoretical result calculated by the Melnikov criterion. One can refer to the memory principle-based method we proposed in another previous work [69] for more details on computing LE of fractional-order systems.

In view of the fact that no further parameter restrictions are introduced in the qualitative analysis, the following given parameter values are used here and serve only for the numerical validation of the Melnikov criterion. They are taken as $m = 1$, $c = 0.4$, $k = 1$, $\beta_1 = 0.1$, $\gamma_1 = -1$ and $\omega = 0.8$, and the simulation conditions are set to $t_0 = 0$, $t_{\text{final}} = 500$ and $h = 0.001$. Numerical validations are performed under different orders that range from $1.1$ to $1.9$ in steps of $0.1$, and the corresponding chaos thresholds are calculated. First, the bifurcation of the displacement response with excitation amplitude is simulated and the results are illustrated in Fig. 3. Further, the largest nontrivial LE is calculated and combined with the power spectrum to determine the reference value of the chaos threshold. For example, the detailed bifurcation diagram at $q = 1.1$ and the corresponding largest nontrivial LE are illustrated in Fig. 4. Also in this case, the phase trajectories and power spectra of the five typical responses: Period-1, Period-2, Period-4, Period-8, and Chaos are shown in Fig. 5, respectively.

Fig. 4 shows the period-doubling route to chaos, before which the Period-1 orbit of the system first experiences symmetry-breaking. As shown in Fig. 5(a), the Period-1 response frequency is the same as the excitation frequency. Then, the increase in excitation amplitude causes the second harmonic to appear in the response as shown in Fig. 5(b), and the symmetry of the orbit is broken due to the presence of this even harmonic. This symmetry-breaking is a precursor to period-doubling bifurcations, as generally only asymmetric orbits can undergo such bifurcations [70-72]. Further increase in excitation amplitude leads to a full period-doubling cascade ultimately reaching chaos. The intrinsic pseudo-randomness of chaotic response manifests itself as a low-power, broad-band response in the power spectrum as shown in Fig. 5(e), and the local divergence property of the chaotic trajectories manifests itself as the presence of at least one positive LE. Based on these two facts, it is determined that the reference value of the chaos threshold at order $q = 1.1$ is $F_1 = 0.45115$, and similarly for other orders. Finally, the numerical result of the chaos threshold and the theoretical one calculated by the Melnikov criterion Eq. (15) are compared in Fig. 6.

Fig. 6 suggests the qualitatively significant agreement between the theoretical and numerical result. In addition, it is suggested in Figs. 3 to 5 that the chaotic response occurs as the increase of the excitation amplitude, which corresponds precisely to the conclusion in Section 3, i.e., strong excitation amplitude can lead to the transverse heteroclinic point so that the system enters chaos.

Based on the above two consistencies, the validity of Melnikov criterion is therefore confirmed. Note that, as described in Eq. (7), the Melnikov function is based on a first-order approximation, which is the main reason for the presence of small quantitative differences in Fig. 6. Nevertheless, the Melnikov criterion described by Eq. (15) is currently the most acceptable result compared to the ones obtained by the both local analytical and numerical methods as it is rigorous, closed, and able to characterize the evolution of the chaos threshold globally.

**Table 1.** Correspondences between the type of response of a 4-D system and the signs of LEs.

| Signs of LEs | Response |
|---|---|
| $(+, +, 0, -)$ | Hyperchaos |
| $(+, 0, -, -)$ | Chaos |
| $(+, 0, 0, -)$ | Chaotic 2-torus |
| $(0, 0, -, -)$ | 2-torus |
| $(0, 0, 0, -)$ | 3-torus |
| $(0, 0, 0, 0)$ | 4-torus |
| $(0, -, -, -)$ | Period |
| $(-, -, -, -)$ | Asymptotically stable equilibrium point |

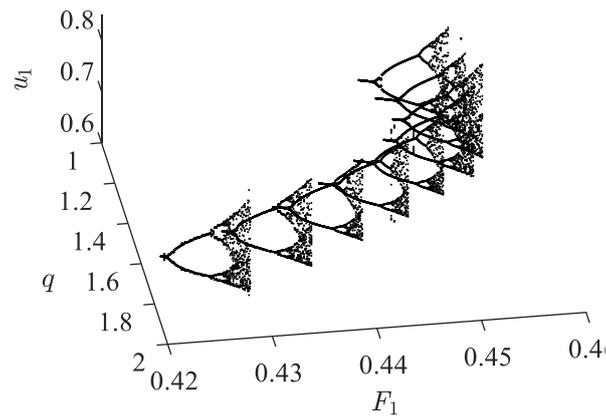

**Fig. 3.** Bifurcation of the perturbed system Eq. (7) under different orders that range from $q = 1.1$ to $1.9$ in steps of $0.1$.

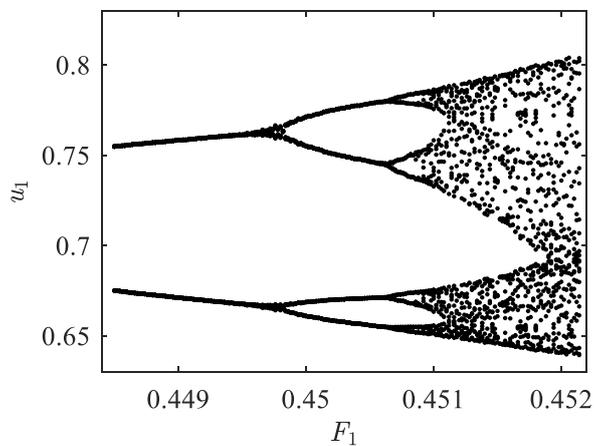

(a)

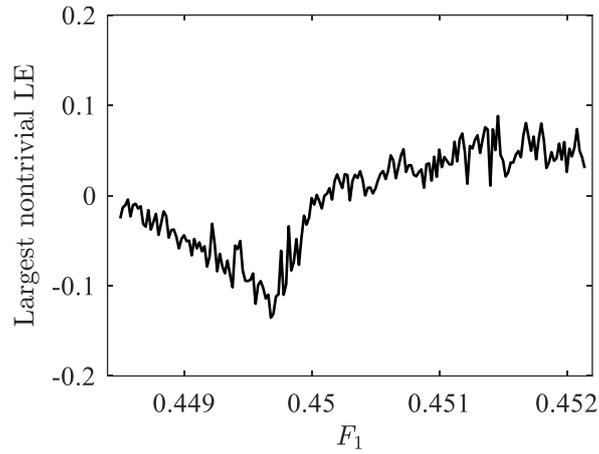

(b)

**Fig. 4.** Dynamics of the perturbed system Eq. (7) at $q = 1.1$. (a) Bifurcation with change in excitation amplitude $F_1$; (b) the corresponding largest nontrivial LE computed by the memory principle-based method [69].

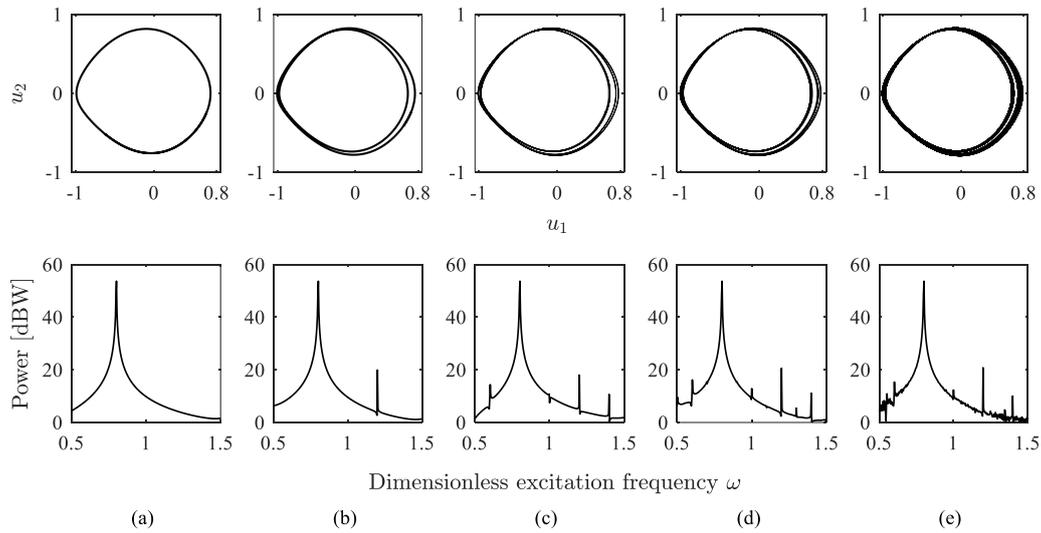

(a)  (b)  (c)  (d)  (e)

**Fig. 5.** Typical responses of the perturbed system Eq. (7) at $q = 1.1$. (a) Period-1 response under $F_1 = 0.4446$; (b) Period-2 response under $F_1 = 0.4493$; (c) Period-4 response under $F_1 = 0.4505$; (d) Period-8 response under $F_1 = 0.4509$; (e) Chaotic response under $F_1 = 0.45115$.

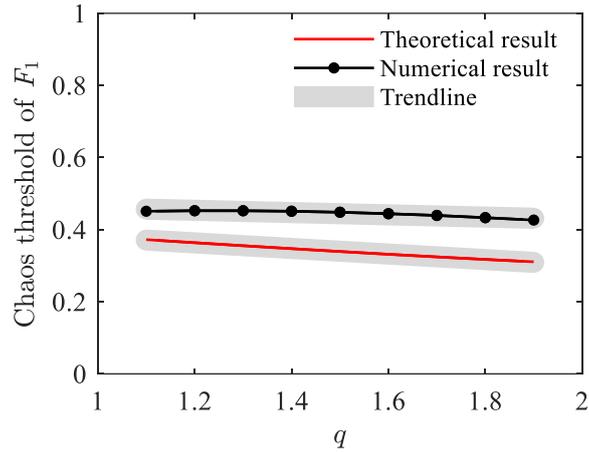

**Fig. 6.** Comparison between the numerical result of the chaos threshold and the corresponding theoretical one calculated by Eq. (15).

## 5 Experimental Study

### 5.1 Experimental setup

In order to validate the Melnikov criterion and route to chaos for granules-beam coupled vibrations revealed by the above theoretical and simulation studies, experimental studies are carried out with the experimental setup shown in Fig. 7.

Monodisperse glass granular media with a diameter of $d_0 = 1.2\,\text{mm}$ is packed in glass containers with sizes of $500 \times 280 \times 275\,\text{mm}$. A slender steel beam with rectangular cross-section is mounted horizontally at the center of $100\,\text{mm}$ from the bottom of the container and is buried in the granular media. The physical and geometric properties of the experimental beam are listed in Table 2. The buried depth $H$ is adjusted by changing the height difference between the top surface of the granular media and the one of the beam. In addition, two steel columns, which are fixed on the foundation, provide the simply-simply supported boundary condition for the beam as illustrated in the insert in Fig. 7. The beam is forced by an external harmonic excitation. The excitation signal is generated by the signal generator (RIGOL, DG-1000) and the power amplifier (Bruel and Kjaer, 2718) and assigned to the exciter (Bruel and Kjaer, 4809) fixed to the foundation, and then the center of the beam is excited by the external harmonic force through a connecting rod. A force sensor (Bruel and Kjaer, 8230-002) is mounted between the connecting rod and the exciter to provide the feedback of excitation force. The response of the beam is measured by three strain gauges (AVIC ZEMIC, BA350-3BB) attached to the center of the bottom surface and 15mm on both sides, and the response data are collected and stored by the dynamic signal testing and analysis system (DongHua, DH-5956).

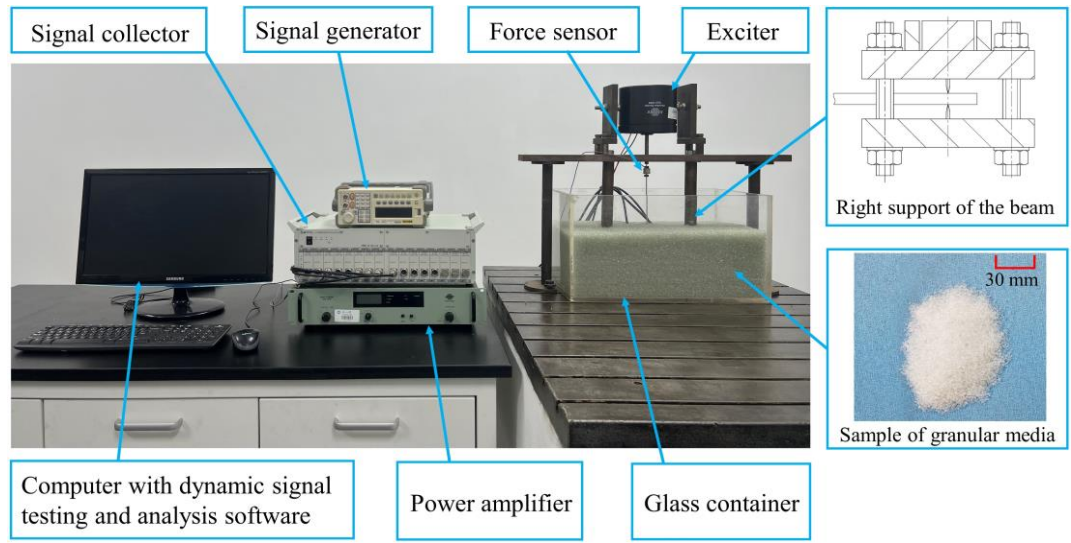

**Fig. 7.** Experimental setup of granules-beam coupled vibration.

**Table 2.** Physical and geometric properties of the experimental beam.

| Item | Symbol | Value |
|---|---|---|
| Length $\times$ Width $\times$ Thickness [$mm^3$] | $L \times b \times h$ | $220 \times 28 \times 3.3$ |
| Young's modulus [Gpa] | $E$ | 210 |
| Mass density [$kg/m^3$] | $\rho$ | 7850 |
| Cross-sectional area [$m^2$] | $S$ | $9.24 \times 10^{-5}$ |
| Second moment of area [$m^4$] | $I$ | $8.3853 \times 10^{-11}$ |

## 5.2 Route to chaos

To validate the route to chaos and Melnikov criterion, experiments are performed at different buried depths $H$, excitation frequencies $f$, and excitation amplitudes $F_0$. The power spectrum and phase trajectory are combined to determine the type of the experimental system response and thus the chaos threshold of the excitation amplitude in the experiments. A typical result obtained at $H = 45\,\mathrm{mm}$ and $f = 350\,\mathrm{Hz}$ is presented in Fig. 8, where the time histories of the deflections at the three measured points on the beam are also shown.

As discussed in Section 3, the single-mode Galerkin discretization is based on the synchronous vibration assumption. Here, the time histories in Fig. 8 clearly show this motion, i.e., all generalized coordinates pass through their amplitude and zero points of potential energy simultaneously, even for the chaotic response as shown in Fig. 8(d). Thus, the reasonability for using the single-mode Galerkin discretization is confirmed.

Importantly, consistent with the conclusion of the Melnikov criterion and numerical simulation, Fig. 8 suggests that the experimental system response enters chaos as the excitation amplitude increases, in which the power spectra show the same period-doubling route to chaos as the simulation results. Specifically, as the excitation amplitude increases, the Period-1 orbit of the system first undergoes symmetry-breaking to the Period-2. Then, the corresponding main peaks representing Period-2 and Period-4 responses appear in the power spectra as shown in Fig. 8(b) and Fig. 8(c). Further increasing the excitation amplitude to the chaos threshold $F_0 = 9.5$ leads to the chaotic response as shown in Fig. 8(d), where the low-power, broad-band response in the power spectrum from $87\,\mathrm{Hz}$ to $260\,\mathrm{Hz}$ reflects the intrinsic pseudo-randomness of the chaotic response. Besides, Similar to the simulation results, the phase trajectories in Fig. 8(d) indicate that the limit state of this chaotic response is a single-scroll attractor, since the system has a saddle-center-saddle equilibrium, and this equilibrium is perturbed by only a single frequency. By contrast, multi-scroll attractors may be exhibited if this equilibrium is perturbed by multiple frequencies [73] or the center-saddle-center equilibrium is perturbed by a single frequency [72].

In summary, the existence of chaotic response of granular-beam coupled vibration is experimentally verified, which is a single-scroll attractor, and the period-doubling route to chaos is revealed. Moreover, these experimental results are so consistent with those based on the theoretical model that the dynamical mechanism obtained from the Melnikov analysis is supported, i.e., the heteroclinic orbit does exist in phase space for such coupled vibration system, and its split stable and unstable manifolds intersect transversely due to perturbations, which in turn develop heteroclinic tangles, eventually leading the system to enter Smale horseshoe chaos.

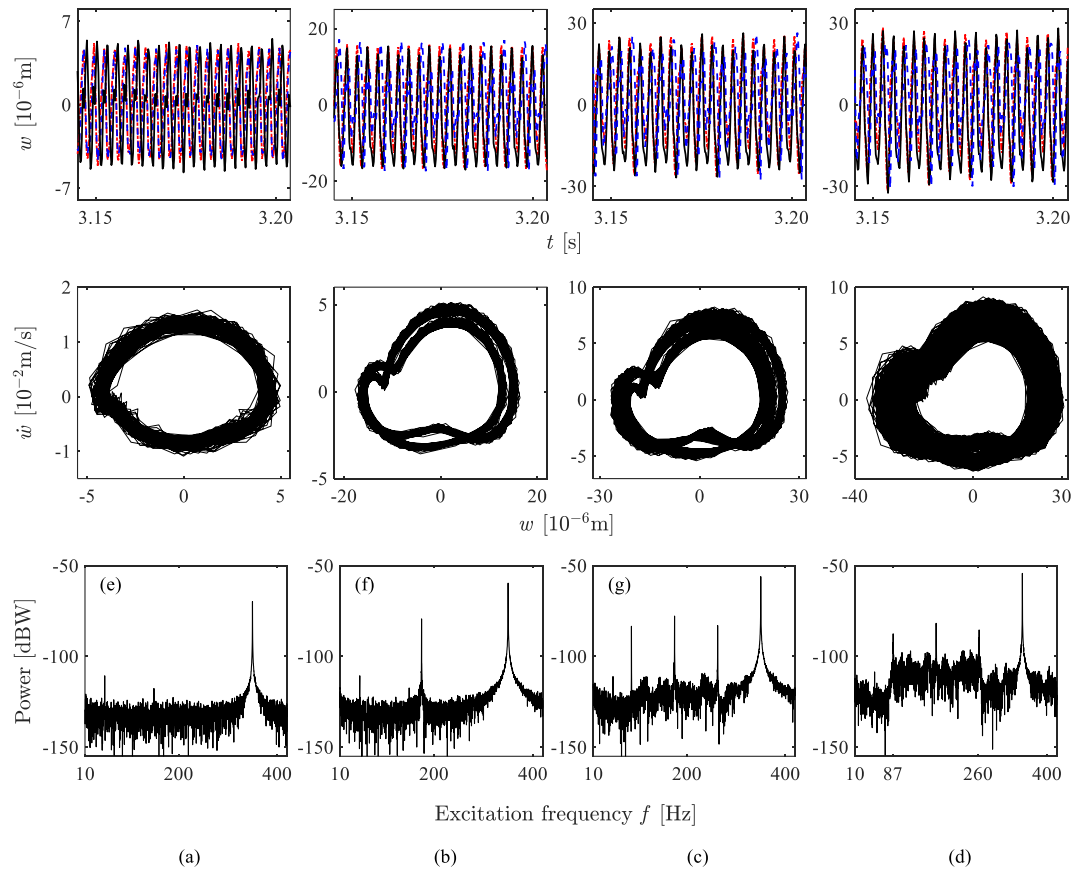

**Fig. 8.** Typical responses of the experimental system at $H = 45\,\text{mm}$ and $f = 350\,\text{Hz}$. The first row shows the time histories of the three measured points (the midpoint and both sides) of the beam; The second row shows the phase trajectories of the response at the midpoint; And the third row shows the power spectra of the response at the midpoint. (a) Period-1 response under $F_0 = 2.5\,\text{N}$; (b) Period-2 response under $F_0 = 6\,\text{N}$; (c) Period-4 response under $F_0 = 9\,\text{N}$; (d) Chaotic response under $F_0 = 9.5\,\text{N}$.

## 5.3 Chaos threshold of experimental system

Following the experimental approach in Section 5.2, it is feasible to determine the chaos thresholds under different experimental conditions and subsequently employ them to qualitatively validate the effectiveness of the Melnikov criterion within the experimental system. First, the chaos threshold under the experimental system parameters is calculated based on the Melnikov criterion.

Determining the four ADL parameters, which can be implemented by the genetic algorithm-based parameter identification procedure introduced in the previous work [1], is required to be performed under periodic vibrations. Nevertheless, the estimated chaos threshold using the parameters identified under periodic motion can still serve as a qualitative reference here. The four ADL parameters are taken as $k_L = 2.3 \times 10^6$ N/m, $k_N = -1.6 \times 10^{15}$ N/m$^3$, $q = 1.4930$ and $\beta_0 = 47.78$ N/$\left( \text{m} \cdot \text{s}^{-q} \right)$, which are identified under the buried depth $H = 45$mm. Also, the first primary resonance, i.e., $n = 1$ in Eq. (5), is considered here for demonstration, and thus the corresponding first order modal damping of the beam is used, which has been experimentally determined [1] to be $c = 5.10912$. The evolution of chaos thresholds on the four parameter planes are computed by the Melnikov criterion, as illustrated in Fig. 9. It is worth stating that all the qualitative trends described in Fig. 9 are general. In other words, changing the values of each parameter only affects the chaos threshold numerically, but not qualitatively. This is the reason why only one set of parameters is used for demonstration here.

As shown in Fig. 9, the fractional-order element that describes the additional inertia-viscous effects provided by the fluidized granular media on the beam exhibit a significant non-monotonic influence on the chaos threshold, which may be related to the competitive mechanism between the equivalent viscous damping and the equivalent inertance. Furthermore, on the whole, the enhancement of the linear stiffness leads to an increase in the chaos threshold, while the enhancement of the nonlinear stiffness leads to a decrease in the chaos threshold. In addition, although in the previous work it has been suggested that the ADL provided by the granular media to the beam is highly complicated, here, it is striking that a general and simple pattern is revealed in all the subplots in Fig. 9, i.e., the chaos threshold of the excitation amplitude is always positively correlated with the excitation frequency. Fortunately, both parameters are external and controllable, so the pattern can be conveniently verified in the experimental system.

To verify the pattern discovered from the Melnikov criterion, the chaos thresholds of the excitation amplitude under different frequencies are obtained through numerous experiments. Meanwhile, to avoid the frequency regions where two periodic attractors of the first primary resonance coexist as well as the frequency region of the second primary resonance, the experimental frequency range is approximately set between them. Furthermore, in order to validate under different ADLs and thus ensure the reliability of the verification, the experiments are performed at three different buried depths, $H = 25\,\text{mm}$, $H = 35\,\text{mm}$, and $H = 45\,\text{mm}$, respectively, and the result is shown in Fig. 10. It is suggested in Fig. 10 that the evolution of the experimental chaos threshold agree well with the pattern revealed by the Melnikov criterion, that is, the chaos threshold of the excitation amplitude is always positively correlated with the excitation frequency, even under different ADLs which are implemented by adjusting the buried depth $H$.

In summary, these good agreements between experiment and theory suggest that the Melnikov criterion can serve to analytically estimate and predict the chaos threshold of the granules-beam coupled vibration.

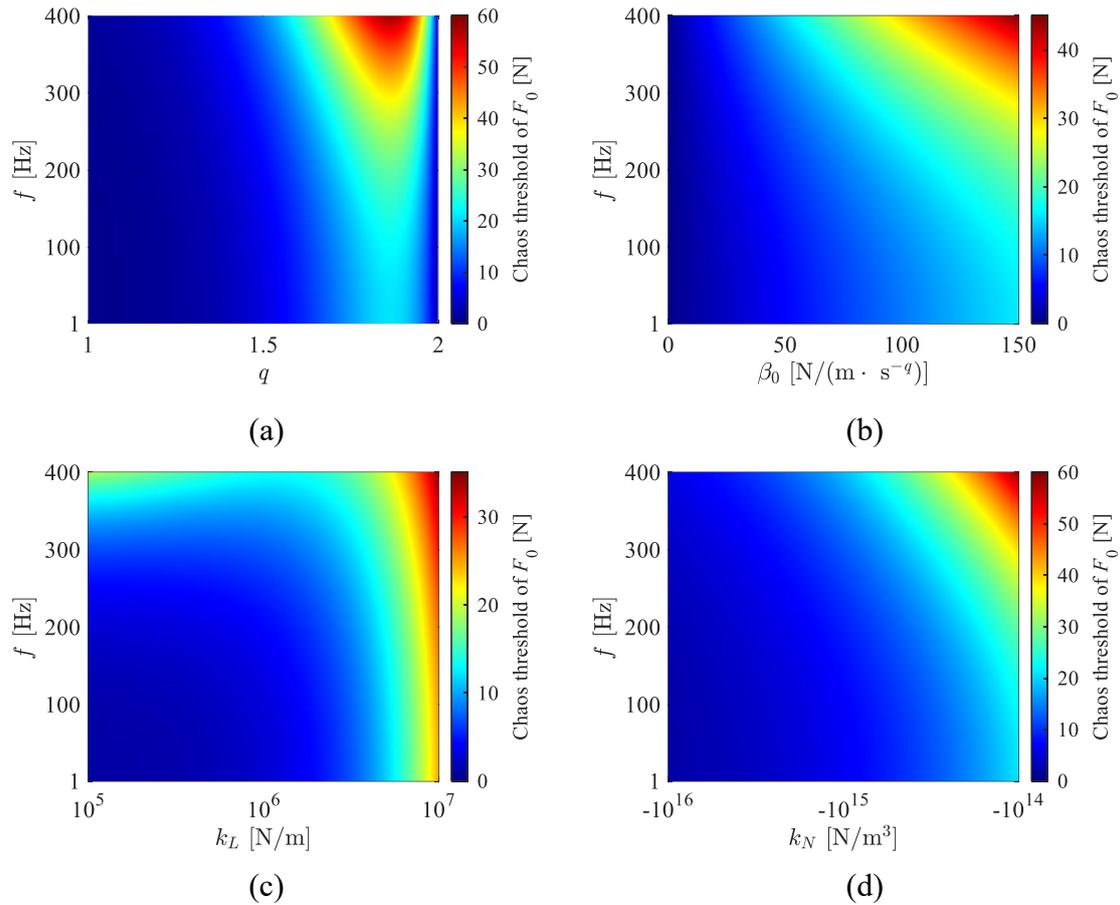

(a)

(b)

(c)

(d)

**Fig. 9.** Evolution of the chaos threshold calculated by the Melnikov criterion, where the four parameter planes are composed of each of the four ADL parameters and the excitation frequency in turn. The chaos thresholds of excitation amplitude are illustrated in different colors. (a) Evolution on the $q-f$ parameter plane; (b) Evolution on the $\beta_0 - f$ parameter plane; (c) Evolution on the $k_L - f$ parameter plane; (d) Evolution on the $k_N - f$ parameter plane.

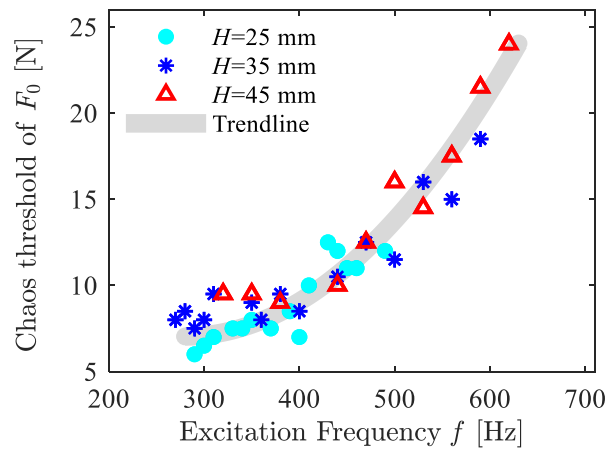

**Fig. 10.** Evolution of the chaos threshold in the experimental system under different ADLs implemented by adjusting the buried depth $H$ . The system responses experienced the same period-doubling bifurcation as in Fig. 8 before reaching these chaos thresholds.

## 6 Conclusions and remarks

In this study, the chaotic dynamics of the granules-beam coupled vibration is revealed by combining qualitative theory and experimental studies. In addition, the problem in the existing research on Melnikov analysis of fractional-order chaotic systems is addressed, and the rigorous analysis result is further provided for reference. Specifically, the main work and contributions can be summarized as follows.

(1) The chaotic dynamics of the granules-beam coupled vibration are analyzed by Melnikov method. The Melnikov criterion of fractional-order Duffing systems is obtained for the first time in a global and closed form, and it is verified numerically. The result not only provides an analytical prediction tool for the chaotic dynamics of granules-beam coupled vibration, but also provides suggestions and references for existing and future studies on chaotic dynamics of fractional-order systems.

(2) It is confirmed by experimental studies that the route to chaos predicted by the theory does exist in the granules-beam coupled vibration. It is found that the system first experiences the symmetry-breaking of periodic orbits, then the complete period-doubling cascade, and finally enters single-scroll chaos.

(3) Based on parametric experiments, the general and simple evolution pattern of the chaos threshold for granules-beam coupled vibration, which is revealed by the Melnikov criterion, is also verified, i.e., regardless of the ADL provided by the granular media on the beam, the chaos threshold of the excitation amplitude is always positively related to the excitation frequency.

In addition to the structural vibration and nonlinear dynamics perspectives focused on in the current work, it is also worth considering that such coupled vibration may also involve wave propagation and nonlinear contact in granular media [74]. Therefore, the investigation on the microscopic behavior of granular media is recommended for future research. For such problems, photoelastic experiments [28] and X-ray tomography [75] may provide insights into the details of granular media, such as the evolution of force chains and spatial structure.

**Declaration of interests**

The authors declare that they have no known competing financial interests or personal relationships that could have appeared to influence the work reported in this paper.

**Data Availability**

All data that support the findings of this work are available from the corresponding author upon reasonable request.

**CRediT authorship contribution statement**

**Hang Li**: Conceptualization, Methodology, Formal analysis, Investigation, Writing – original draft. **Jian Li**: Conceptualization, Funding acquisition, Supervision, Project administration, Writing – review & editing. **Hongzhu Fei**: Data curation, Software. **Guangyang Hong**: Funding acquisition, Visualization. **Jinlu Dong**: Validation, Formal analysis. **Aibing Yu**: Supervision, Resources.

**Acknowledgments**

This work is supported by the National Natural Science Foundation of China (Grant Nos. 12272091 and 12302510), China Postdoctoral Science Foundation (Certificate Number: 2023M740549), and the Fundamental Research Funds for the Central Universities (No. N2305015). Hang Li would like to acknowledge, in particular, Mr. Nathan from University of Oxford for his great help and insightful discussions in complex analysis on Appendix A; Prof. Yongjun Shen from Shijiazhuang Tiedao University for his invaluable guidance on fractional dynamics and chaos theory; Prof. Lu Liu from Dalian University of Technology for his great help in the experimental study; and Prof. Shu Zhang from Tongji University for his insightful discussion on the theory of nonlinear vibration.

## Appendix A

The integral $I_3$ produced from Eq. (12) is calculated as follows.

$$I_3 = \int_{-\infty}^{+\infty} \mathrm{sech}^2 \frac{\omega_0 t}{\sqrt{2}} \int_{-\infty}^{t} (t-s)^{1-q} \tanh \frac{\omega_0 s}{\sqrt{2}} \mathrm{sech}^2 \frac{\omega_0 s}{\sqrt{2}} \mathrm{d}s \, \mathrm{d}t \tag{A1}$$

First, to evaluate the inner integral $J(t)$ in the complex plane, it is defined that

$$\psi(z) = \mathrm{e}^{p\ln(t-z)} \tanh(\varphi z) \mathrm{sech}^2(\varphi z) \tag{A2}$$

where the parameters $p = 1-q$ and $\varphi = \dfrac{\omega_0}{\sqrt{2}}$. The function $\psi : \Psi \to \mathbb{C}$, and $\Psi$ is defined as $\mathbb{C}$ takeout $t$ and all zeroes of $\cosh(\varphi z)$. Hence, $\psi$ is holomorphic away from $(-\infty, t]$. Then, the residue theorem can be applied to integrate $\psi(z)$ as follows.

$$\int_{\mathscr{H}} \psi(z) = 2\pi\mathrm{i} \sum_{n\in\mathbb{Z}} \mathrm{Res}\left(\psi; \frac{\pi\mathrm{i}}{2\varphi}(2n+1)\right) \tag{A3}$$

where $\mathscr{H}$ stands for a Hankel contour that winds $-\infty + \mathrm{i}0^+ \to t + \mathrm{i}0^+ \to t + \mathrm{i}0^- \to -\infty + \mathrm{i}0^-$.

Note that there is also equivalently:

$$J(t) = \frac{\mathrm{i}\mathrm{e}^{-ip\pi}}{2} \csc(p\pi) \cdot \int_{\mathscr{H}} \psi(z) \tag{A4}$$

Then, by combining Eq. (A3) and Eq. (A4) and calculating the residues in Eq. (A3), the inner integral $J(t)$ is obtained as follows.

$$J(t) = -\frac{\pi q(q-1)}{\varphi^3} \csc(q\pi) \cdot \Re\left( \sum_{n\in\mathbb{N}} \frac{e^{p\ln\left(\frac{\pi i}{2\varphi}(2n+1)-t\right)}}{\left(t - \frac{\pi i}{2\varphi}(2n+1)\right)^2} \right) \tag{A5}$$

where $\Re(\bullet)$ represents the sum of a term and its symmetric one regarding $i$.

Further, $I_3$ continues to be evaluated as follows.

$$I_3 = \frac{-\pi q(q-1)\cdot\csc(q\pi)}{\varphi^{3-q}} \cdot \Re\left( \overbrace{\int_{-\infty}^{+\infty} \underbrace{\mathrm{sech}^2(t)\cdot\left(\sum_{n\in\mathbb{N}} e^{-(q+1)\ln\left(\frac{\pi i}{2}(2n+1)-t\right)}\right)}_{\Lambda(t)}\mathrm{d}t}^{I_4} \right) \tag{A6}$$

where $I_4$ can then be integrated by applying the residue theorem again as follows.

$$I_4 = -2\pi i \cdot \sum_{\lambda\in\mathbb{N}} \mathrm{Res}\left( \Lambda; -\frac{\pi i}{2}(2\lambda+1) \right) \tag{A7}$$

By substituting Eq. (A7) into Eq. (A6) and calculating the residues, $I_3$ is finally obtained in the following exact and closed-form.

$$I_3 = 2^{\frac{3-q}{2}} \cdot \frac{q(q^2-1)}{\pi^q \omega_0^{3-q}} \cdot \sec\frac{q\pi}{2} \cdot \zeta(q+1) \tag{A8}$$

where $\zeta(\bullet)$ stand for Riemann Zeta function.

On the parameter plane composed of the only two parameters $q$ and $\omega_0$ involved in $I_3$, it is easy to verify the analytical result by numerical integration, as shown in Fig. A1. It is suggested in the Fig. A1 that the analytical results and the numerical integration match excellently in quantity.

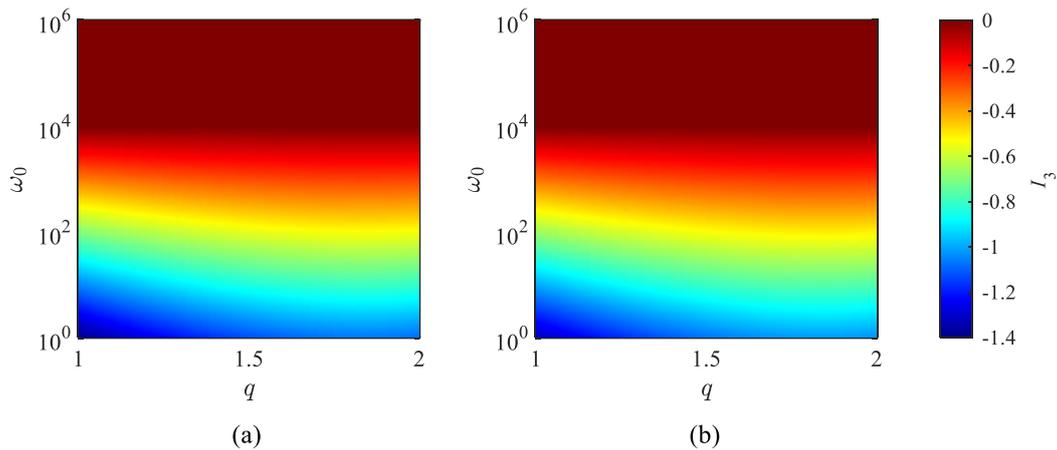

**Fig. A1.** Integration results for $I_3$ on the $q - \omega_0$ parameter plane, where the values integrated are illustrated in different colors. (a) Numerical integration by Eq. (A1); (b) Analytical result by Eq. (A8).

# References

[1] H. Li, J. Li, G. Hong, et al. Fractional-order model and experimental verification of granules-beam coupled vibration. Mechanical Systems and Signal Processing, 2023, 200: 110536. https://doi.org/10.1016/j.ymssp.2023.110536

[2] J. W. S. B. Rayleigh. The theory of sound, 2nd edition. Macmillan, 1896.

[3] H. Lamb. On the vibrations of an elastic plate in contact with water. Proceedings of the Royal Society of London, Series A, Mathematical, Physical and Engineering sciences, 1920, 98(690): 205-216. https://doi.org/10.1098/rspa.1920.0064

[4] G. G. Stokes, W. Thomson. The Correspondence between Sir George Gabriel Stokes and Sir William Thomson, Baron Kelvin of Largs. Cambridge University Press, 1990. https://cambridge.org/9780521328319

[5] J. E. Sader. Frequency response of cantilever beams immersed in viscous fluids with applications to the atomic force microscope. Journal of Applied Physics, 1998, 84(1): 64-76. https://doi.org/10.1063/1.368002

[6] Q. Ni, M. Li, M. Tang, et al. Free vibration and stability of a cantilever beam attached to an axially moving base immersed in fluid. Journal of Sound and Vibration, 2014, 333(9): 2543-2555. https://doi.org/10.1016/j.jsv.2013.11.049

[7] X. Fang, J. Lou, J. Huang, et al. Hydrodynamic effect and Fluid-Structure coupled

vibration of underwater flexible caudal fin actuated by Macro fiber composites. Mechanical Systems and Signal Processing, 2023, 192: 110233. https://doi.org/10.1016/j.ymssp.2023.110233

[8] Y. Q. Wang, X. B. Huang, J. Li. Hydroelastic dynamic analysis of axially moving plates in continuous hot-dip galvanizing process. International Journal of Mechanical Sciences, 2016, 110: 201-216. https://doi.org/10.1016/j.ijmecsci.2016.03.010

[9] Y. Q. Wang, H. Wu, F. Yang, et al. An efficient method for vibration and stability analysis of rectangular plates axially moving in fluid. Applied Mathematics and Mechanics, 2021, 42(2): 291-308. https://doi.org/10.1007/s10483-021-2701-5

[10] Y. Q. Wang, Z. Yang. Nonlinear vibrations of moving functionally graded plates containing porosities and contacting with liquid: internal resonance. Nonlinear Dynamics, 2017, 90: 1461-1480. https://doi.org/10.1007/s11071-017-3739-z

[11] X. Y. Mao, J. Jing, H. Ding, et al. Dynamics of axially functionally graded pipes conveying fluid. Nonlinear Dynamics, 2023, 111: 11023-11044. https://doi.org/10.1007/s11071-023-08470-2

[12] J. R. Yuan, H. Ding. Three-dimensional dynamic model of the curved pipe based on the absolute nodal coordinate formulation. Mechanical Systems and Signal Processing, 2023, 194: 110275. https://doi.org/10.1016/j.ymssp.2023.110275

[13] H. Ding, J. Ji, L. Q. Chen. Nonlinear vibration isolation for fluid-conveying pipes using quasi-zero stiffness characteristics. Mechanical Systems and Signal Processing, 2019, 121: 675-688. https://doi.org/10.1016/j.ymssp.2018.11.057

[14] P. Richard, M. Nicodemi, R. Delannay, et al. Slow relaxation and compaction of granular systems. Nature Materials, 2005, 4(2): 121-128. https://doi.org/10.1038/nmat1300

[15] H. Askari, K. Kamrin. Intrusion rheology in grains and other flowable materials. Nature materials, 2016, 15(12): 1274-1279. https://doi.org/10.1038/nmat4727

[16] A. R. Mojdehi, B. Tavakol, W. Royston, et al. Buckling of elastic beams embedded in granular media. Extreme Mechanics Letters, 2016, 9: 237-244. https://doi.org/10.1016/j.eml.2016.03.022

[17] G. Jing, P. Aela, H. Fu. The contribution of ballast layer components to the lateral

resistance of ladder sleeper track. Construction and Building Materials, 2019, 202: 796-805. https://doi.org/10.1016/j.conbuildmat.2019.01.017

[18] G. Hong, J. Li, J. Dong, et al. Frequency-dependent characteristics of grain-beam system: Negative mass and jump behaviour. International Journal of Mechanical Sciences, 2021, 209: 106706. https://doi.org/10.1016/j.ijmecsci.2021.106706

[19] J. Pan, J. Li, G. Hong, et al. A mapping discrete element method for nonlinear dynamics of vibrating plate-particle coupling system. Powder Technology, 2021, 385: 478-489. https://doi.org/10.1016/j.powtec.2021.03.022

[20] J. Dong, J. Fang, J. Pan, et al. Dynamic model of vibrating plate coupled with a granule bed. Chaos, Solitons & Fractals, 2022, 156: 111857. https://doi.org/10.1016/j.chaos.2022.111857

[21] Y. Ning, G. Hong, J. Li, et al. Theoretical and experimental investigation on nonlinear dynamic of grain-beam system. International Journal of Mechanical Sciences, 2023: 108751. https://doi.org/10.1016/j.ijmecsci.2023.108751

[22] J. Dong, Z. Niu, J. Li, et al. System identification of excited beam immersed in granular materials: A multifrequency data-based method in a variational framework. Engineering Structures, 2023, 293: 116611. https://doi.org/10.1016/j.engstruct.2023.116611

[23] X. Long, L. Liu, S. Ji. Discrete Element Analysis of Ice-Induced Vibrations of Offshore Wind Turbines in Level Ice. Journal of Marine Science and Engineering, 2023, 11(11): 2153. https://doi.org/10.3390/jmse11112153

[24] L. Liu, J. Li, C. Wan. Nonlinear dynamics of excited plate immersed in granular matter. Nonlinear Dynamics, 2018, 91: 147-156. https://doi.org/10.1007/s11071-017-3861-y

[25] Y. Wang, L. Li, D. Hofmann, et al. Structured fabrics with tunable mechanical properties. Nature, 2021, 596(7871): 238-243. https://doi.org/10.1038/s41586-021-03698-7

[26] Y. Yan, S. Ji. Discrete element modeling of direct shear tests for a granular material. International Journal for Numerical and Analytical Methods in Geomechanics, 2010, 34(9): 978-990. https://doi.org/10.1002/nag.848

[27] H. M. Jaeger, S. R. Nagel, R. P. Behringer. Granular solids, liquids, and gases.

Reviews of Modern Physics, 1996, 68(4): 1259. https://doi.org/10.1103/RevModPhys.68.1259

[28] G. D. Acar, P. Ravula, B. Balachandran. Dynamic interactions of a driven pendulum with photoelastic granular media. Physics Letters A, 2021, 396: 127244. https://doi.org/10.1016/j.physleta.2021.127244

[29] J. Bai, J. Li, G. Hong, et al. Mesoscopic evolution and kinetic properties of dense granular flow crystallization under continuous shear induction. Powder Technology, 2023, 426: 118615. https://doi.org/10.1016/j.powtec.2023.118615

[30] G. Hong, Y. Zhou, J. Li. Relaxation dynamics of vibrated dense granular media: Hysteresis and nonlocal effects. Powder Technology, 2022, 410: 117847. https://doi.org/10.1016/j.powtec.2022.117847

[31] S. Ji, S. Wang, Z. Peng. Influence of external pressure on granular flow in a cylindrical silo based on discrete element method. Powder Technology, 2019, 356: 702-714. https://doi.org/10.1016/j.powtec.2019.08.083

[32] P. C. Johnson, R. Jackson. Frictional-collisional constitutive relations for granular materials, with application to plane shearing. Journal of Fluid Mechanics, 1987, 176: 67-93. https://doi.org/10.1017/S0022112087000570

[33] P. Jop, Y. Forterre, O. Pouliquen. A constitutive law for dense granular flows. Nature, 2006, 441(7094): 727-730. https://doi.org/10.1038/nature04801

[34] X. Feng, X. Jing. Human body inspired vibration isolation: beneficial nonlinear stiffness, nonlinear damping & nonlinear inertia. Mechanical Systems and Signal Processing, 2019, 117: 786-812. https://doi.org/10.1016/j.ymssp.2018.08.040

[35] Z. Zhu, Y. Wang, Y. Wang, et al. Nonlinear inertia and its effect within an X-shaped mechanism - Part I: Modelling & nonlinear properties. Mechanical Systems and Signal Processing, 2023, 200: 110590. https://doi.org/10.1016/j.ymssp.2023.110590

[36] X. Jing, Z. Zhu, Y. Guo, et al. Nonlinear inertia and its effect within an X-shaped mechanism - Part II: Nonlinear influences and experimental validations. Mechanical Systems and Signal Processing, 2023, 200: 110591. https://doi.org/10.1016/j.ymssp.2023.110591

[37] Y. Shen, P. Sui, X. Wang. Performance analysis and optimization of bimodal

nonlinear energy sink. Nonlinear Dynamics, 2023, 111(18): 16813-16830. https://doi.org/10.1007/s11071-023-08737-8

[38] P. Sui, Y. Shen, X. Wang. Study on response mechanism of nonlinear energy sink with inerter and grounded stiffness. Nonlinear Dynamics, 2023, 111(8): 7157-7179. https://doi.org/10.1007/s11071-022-08226-4

[39] Y. Shen, P. Sui. Dynamics analysis and parameter optimization of a vibration absorber with geometrically nonlinear inerters. Journal of Vibration and Control, 2023: 10775463231217532. https://doi.org/10.1177/10775463231217532

[40] Z. Xing, X. Yang. A combined vibration isolation system capable of isolating large amplitude excitation. Nonlinear Dynamics, 2023, 112: 2523-2544 https://doi.org/10.1007/s11071-023-09166-3

[41] H. Y. Chen, X. Y. Mao, H. Ding, et al. Elimination of multimode resonances of composite plate by inertial nonlinear energy sinks. Mechanical Systems and Signal Processing, 2020, 135: 106383. https://doi.org/10.1016/j.ymssp.2019.106383

[42] Z. Zhang, Z. Q. Lu, H. Ding, et al. An inertial nonlinear energy sink. Journal of Sound and Vibration, 2019, 450: 199-213. https://doi.org/10.1016/j.jsv.2019.03.014

[43] Y. Shen, Z. Xing, S. Yang, et al. Parameters optimization for a novel dynamic vibration absorber. Mechanical Systems and Signal Processing, 2019, 133: 106282. https://doi.org/10.1016/j.ymssp.2019.106282

[44] H. Shi, S. Fan, W. Xing, et al. Study of weak vibrating signal detection based on chaotic oscillator in MEMS resonant beam sensor. Mechanical Systems and Signal Processing, 2015, 50: 535-547. https://doi.org/10.1016/j.ymssp.2014.05.015

[45] J. Wu, Y. Wang, W. Zhang, et al. Defect detection of pipes using Lyapunov dimension of Duffing oscillator based on ultrasonic guided waves. Mechanical Systems and Signal Processing, 2017, 82: 130-147. https://doi.org/10.1016/j.ymssp.2016.05.012

[46] C. Li, L. Qu. Applications of chaotic oscillator in machinery fault diagnosis. Mechanical Systems and Signal Processing, 2007, 21: 257-269. https://doi.org/10.1016/j.ymssp.2005.07.006

[47] X. Ma, H. Li, S. Zhou, et al. Characterizing nonlinear characteristics of asymmetric

tristable energy harvesters. Mechanical Systems and Signal Processing, 2022, 168: 108612. https://doi.org/10.1016/j.ymssp.2021.108612

[48] Y. Wang, F. Li, H. Shu. Nonlocal nonlinear chaotic and homoclinic analysis of double layered forced viscoelastic nanoplates. Mechanical Systems and Signal Processing, 2019, 122: 537-554. https://doi.org/10.1016/j.ymssp.2018.12.041

[49] V. K. Melnikov. On the stability of a center for time-periodic perturbations. Tr. Mosk. Mat. Obs., 1963, 12: 3-52, (in Russian). https://www.mathnet.ru/eng/mmo137

[50] G. M. Zaslavsky, M. Edelman, B. A. Niyazov. Self-similarity, renormalization, and phase space nonuniformity of Hamiltonian chaotic dynamics. Chaos, 1997, 7: 159-181. https://doi.org/10.1063/1.166252

[51] J. Niu, R. Liu, Y. Shen, et al. Chaos detection of Duffing system with fractional-order derivative by Melnikov method. Chaos, 2019, 29: 123106. https://doi.org/10.1063/1.5124367

[52] S. Wen, H. Qin, Y. Shen, et al. Chaos threshold analysis of Duffing oscillator with fractional-order delayed feedback control. The European Physical Journal Special Topics, 2022, 231(11): 2183-2197. https://doi.org/10.1140/epjs/s11734-021-00369-6

[53] Y. Zhang, J. Li, S. Zhu, et al. Bifurcation and chaos detection of a fractional Duffing-van der Pol oscillator with two periodic excitations and distributed time delay. Chaos, 2023, 33: 083153. https://doi.org/10.1063/5.0160812

[54] R. Li, Q. Li, D. Huang. Taming chaos in generalized Lienard systems by the fractional-order feedback based on Melnikov analysis. Physica Scripta, 2023, 98(8): 085214. https://doi.org/10.1088/1402-4896/ace28d

[55] J. Xie, R. Guo, Z. Ren, et al. Vibration resonance and fork bifurcation of under-damped Duffing system with fractional and linear delay terms. Nonlinear Dynamics, 2023, 111: 10981-10999. https://doi.org/10.1007/s11071-023-08462-2

[56] J. Zhang, J. Xie, W. Shi, et al. Resonance and bifurcation of fractional quintic Mathieu-Duffing system. Chaos, 2023, 33: 023131. https://doi.org/10.1063/5.0138864

[57] J. Xie, M. Wan, F. Zhao, et al. Dynamic perturbation analysis of fractional order

differential quasiperiodic Mathieu equation. Chaos, 2023, 33: 123118. https://doi.org/10.1063/5.0163991

[58] G. T. Oumbe Tekam, C. A. Kitio Kwuimy, P. Woafo. Analysis of tristable energy harvesting system having fractional order viscoelastic material. Chaos, 2015, 25: 013112. https://doi.org/10.1063/1.4905276

[59] L.M. Anague Tabejieu, B.R. Nana Nbendjo, P. Woafo. On the dynamics of Rayleigh beams resting on fractional-order viscoelastic Pasternak foundations subjected to moving loads. Chaos, Solitons & Fractals, 2016, 93: 39-47. https://doi.org/10.1016/j.chaos.2016.10.001

[60] A. M. Ngounou, S. C. Mba Feulefack, L. M. Anague Tabejieu, et al. Design, analysis and horseshoes chaos control on tension leg platform system with fractional nonlinear viscoelastic tendon force under regular sea wave excitation. Chaos, Solitons & Fractals, 2022, 157: 111952. https://doi.org/10.1016/j.chaos.2022.111952

[61] M. Wang, W. Ma, E. Chen, et al. Confusion threshold study of the Duffing oscillator with a nonlinear fractional damping term. Journal of Low Frequency Noise, Vibration and Active Control, 2021, 40(2): 929-947. https://doi.org/10.1177/1461348420922686

[62] E. Chen, W. Xing, M. Wang, et al. Study on chaos of nonlinear suspension system with fractional-order derivative under random excitation. Chaos, Solitons & Fractals, 2021, 152: 111300. https://doi.org/10.1016/j.chaos.2021.111300

[63] C. Wang, M. Wang, W. Xing, et al. Bifurcation and Chaotic Behavior of Duffing System with Fractional-Order Derivative and Time Delay. Fractal and Fractional, 2023, 7(8): 638. https://doi.org/10.3390/fractalfract7080638

[64] C. Li, F. Zeng. Numerical methods for fractional calculus. CRC Press, 2015.

[65] A. H. Nayfeh, D. T. Mook. Nonlinear oscillations. Wiley, 2008.

[66] G. Hong, J. Bai, J. Li, et al. Unjamming and yielding of intruder-deformation-driven dense granular materials. Powder Technology, 2023, 428: 118784. https://doi.org/10.1016/j.powtec.2023.118784

[67] S. Wiggins. Global bifurcations and chaos: analytical methods. Springer, 2013. https://doi.org/10.1007/978-1-4612-1042-9

[68] Y. Shen, S. Yang, H. Xing, et al. Primary resonance of Duffing oscillator with fractional-order derivative. Communications in Nonlinear Science and Numerical Simulation, 2012, 17(7): 3092-3100. https://doi.org/10.1016/j.cnsns.2011.11.024

[69] H. Li, Y. Shen, Y. Han, et al. Determining Lyapunov exponents of fractional-order systems: A general method based on memory principle. Chaos, Solitons & Fractals, 2023, 168: 113167. https://doi.org/10.1016/j.chaos.2023.113167

[70] D. D'Humieres, M. R. Beasley, B. A. Huberman, et al. Chaotic states and routes to chaos in the forced pendulum. Physical Review A, 1982, 26(6): 3483. https://doi.org/10.1103/PhysRevA.26.3483

[71] A. H. Nayfeh, B. Balachandran. Applied nonlinear dynamics: analytical, computational, and experimental methods. Wiley, 2008.

[72] I. Kovacic, M. J. Brennan. The Duffing equation: nonlinear oscillators and their behaviour. Wiley, 2011.

[73] H. Li, Y. Shen, S. Yang, et al. Simultaneous primary and super-harmonic resonance of Duffing oscillator. Acta Physica Sinica, 2021, 70 (4): 040502 (in Chinese). https://doi.org/10.7498/aps.70.20201059

[74] G. D. Acar, B. Balachandran. Dynamics of one-dimensional granular arrays with pre-compression. Nonlinear Dynamics, 2020, 99: 707-720. https://doi.org/10.1007/s11071-019-05407-6

[75] Y. Xing, Y. Yuan, H. Yuan, et al. Origin of the critical state in sheared granular materials. Nature Physics, 2024, 20: 646-652. https://doi.org/10.1038/s41567-023-02353-4